\documentclass[aps,prd,showpacs,floatfix,nofootinbib,superscriptaddress,twocolumn,10pt]{revtex4}
\usepackage{graphicx}
\usepackage{bm}
\usepackage{subfigure}
\usepackage{amssymb}
\usepackage{color,soul}
\usepackage{graphicx}
\usepackage{multirow}
\graphicspath{{figure/}}
\DeclareGraphicsExtensions{.pdf,.png,.jpg}
\setulcolor{red}

\newcommand{\beq}{\begin{equation}}
\newcommand{\eeq}{\end{equation}}
\newcommand{\beqa}{\begin{eqnarray}}
\newcommand{\eeqa}{\end{eqnarray}}

\begin{document}

\title{Mirror Dark Matter and Electronic Recoil Events in XENON1T}

\author{Lei Zu}
\affiliation{Key Laboratory of Dark Matter and Space Astronomy, Purple Mountain Observatory, Chinese Academy of Sciences, Nanjing 210033, China}
\affiliation{School of Astronomy and Space Science, University of Science and Technology of China, Hefei, Anhui 230026, China}

\author{Guan-Wen Yuan}
\affiliation{Key Laboratory of Dark Matter and Space Astronomy, Purple Mountain Observatory, Chinese Academy of Sciences, Nanjing 210033, China}
\affiliation{School of Astronomy and Space Science, University of Science and Technology of China, Hefei, Anhui 230026, China}

\author{Lei Feng\footnote{Corresponding author: fenglei@pmo.ac.cn}}

\affiliation{Key Laboratory of Dark Matter and Space Astronomy, Purple Mountain Observatory, Chinese Academy of Sciences, Nanjing 210023, China}
\affiliation{Joint Center for Particle, Nuclear Physics and Cosmology,  Nanjing University -- Purple Mountain Observatory,  Nanjing  210093, China}

\author{Yi-Zhong Fan\footnote{Corresponding author: yzfan@pmo.ac.cn}}
\affiliation{School of Astronomy and Space Science, University of Science and Technology of China, Hefei, Anhui 230026, China}
\affiliation{Key Laboratory of Dark Matter and Space Astronomy, Purple Mountain Observatory, Chinese Academy of Sciences, Nanjing 210023, China}

\begin{abstract}
Recently, the XENON1T experiment has reported the possible detection of an excess in the electronic recoil spectrum. Such an excess may indicate the presence of new physics. In this work, we suggest that the scattering of mirror electrons with ordinary electrons through photon$-$mirror photon kinetic mixing with parameter $\epsilon \sim 10^{-12}(n_{\rm e'}/0.2{\rm cm^{-3}})^{-1/2}({v_{\rm c}^0/5\times 10^{9}~{\rm cm~s^{-1}}})^{1/2}$ may account for the excess electronic recoil events in XENON1T, where $n_{\rm e'}$ is the density of mirror electron and $v_{\rm c}^0$ is the cutoff velocity of the mirror electron arriving at the earth. Interestingly, this parameter to interpret the excess of XENON1T electronic recoil spectrum are consistent with the constrains of Darkside50.

\end{abstract}

\maketitle

\section{Introduction}

The existence of dark matter is one of the most mysterious question in modern physics. Many experiments, either direct or indirect, have been carried out to search for the signals of dark matter particles. Recently, the electronic recoil spectrum of the XENON1T experiment has been reported \cite{Xenon1t} with an unexpected excess at low energy. Though the significance of this signal is relatively low and the contribution of the tritium background should be better understood, the possible excess has attracted wide attention. This is because such an excess, if confirmed, would indicate the presence of new physics, as widely discussed in the literature \cite{1,2,3,4,5,6,7,8,9,10,11,12,13,14,15,16,17,18,19,20,21,boost,23,24,25,26,27}. Many models, including for example dark matter, neutrino magnetic moment, tritium components et.al\cite{Xenon1t,1,2,3,4,5,6,7,8,9,10,11,12,13,14,15,16,17,18,19,20,21,boost,23,24,25,26,27,28,29,solarre}, could provide a possible explaination for this excess.

In the mirror dark matter model, each of the known particles has a mirror partner and it interacts with ordinary particles through photon$-$mirror photon kinetic mixing \cite{ask1,ask2,ask3,ask4,ask5,ask6,ask7,ask8,LHC,review}.
Among various candidate particles for dark matter, the mirror dark matter is interesting and can be robustly confirmed or ruled out by the XENON1T experiment \cite{prediction}. In this brief work we examine whether the mirror dark matter can account for the excess of low-energy electronic recoil spectrum or not.

Following \cite{review}, we assume that the hidden sector of mirror dark matter is exact the copy of the ordinary matter sector, so that the Lagrangian is
\begin{equation}
L=L_{SM}(e,\mu,d,\gamma,...)+L_{SM}(e',\mu',d',\gamma',...)+L_{mix},
\label{lagrangian}
\end{equation}
where the particle denoted with $\prime$ represents the corresponding mirror particle. Such a theory
is motivated from the symmetry of left and right handed chiral fields \cite{review}. The allowed interaction terms, which  are consistent with renormalizability and the symmetries of the theory, are $U(1)$ kinetic mixing interaction and Higgs$-$mirror Higgs quartic coupling \cite{higgs,u1}
\begin{equation}
L_{mix}=\frac{\epsilon }{2} F^{\mu \nu}F^{\prime}_{\mu \nu}+\lambda\Phi^{\dagger}\Phi\Phi^{\prime\dagger}\Phi^{\prime},
\label{lmix}
\end{equation}
where $F^{\mu\nu}$ ($F^{\prime}_{\mu \nu}$) is the ordinary (mirror) $U(1)$ gauge boson field strength tensor and $\Phi$ ($\Phi^{\prime}$) is the Higgs (mirror Higgs) field. In this work, we only consider the kinetic mixing of the photon and mirror photon as the interaction between ordinary and mirror particles because of the low energy scale compared to the energy of Higgs field. The astronomical, cosmological observations and the direct detection experiments based on nuclear recoil support the kinetic mixing on the scale about $\epsilon - 10^{-10} \sim 10^{-9}$ (see \cite{review} for a review). The direct detection experiment Darkside50 provided an upper limit about $\epsilon \alpha \leq 1.5 \times 10^{-11}
$ where $\alpha = (n_{\rm e'}/0.2{\rm cm^{-3}})^{1/2}({v_{\rm c}^0/5\times 10^{9}~{\rm cm~s^{-1}}})^{-1/2}$\cite{darkside50}. And we will show in the following that electron-mirror electron scattering with the kinetic mixing  $\epsilon \alpha \sim 10^{-12}$, which is below the limits of Darkside50, can yield a low energy excess in the electronic recoil spectrum of XENON1T. With $\epsilon $ in the range about $ 10^{-10} \sim 10^{-9}$, it suggests that either $v_c^0\gg 50000km/s$ and/or $n_{e'} \ll 0.2cm^{-3}$  \cite{shield}.

This work is structured as the following. In Section II we calculate the electron-mirror electron scattering. We then examine the role of mirror dark matter in shaping the electronic recoil spectrum of XENON1T in Section III. We summarize our results in Section IV.

\section{electron-mirror electron scattering}


To explain the rotation curves in spiral galaxies, some previous studies have shown that the mirror dark matter form a self-interacting spherically
distributed plasma with the temperature of \cite{plasma,review,cdms}
\begin{equation}
T \approx \frac{1}{2} \bar{m} v^{2}_{rot},
\label{temperature}
\end{equation}
where $\bar{m}=\sum n_{i}m_{i}/ \sum n_{i}$ (i=$e',H',He' ...$) is the mean mass of the particles in the plasma and $v_{rot}$ is the galactic rotational velocity ($\sim 220km/s$ for the Milky Way \cite{plasma,review,milkyway}. Thus the gas of mirror electron will have a Maxwellian velocity distribution
\begin{equation}
f_{e}(v)=e^{-\frac{1}{2}m_{e}v^2/T}.
\label{maxwell}
\end{equation}
Assuming that the dark matter density is $\rho=0.3 \rm{GeV}/cm^3$ at the earth’s location and is dominated by the $H', He'$ component, then the number density of the totally ionized  $e'$, $n_{e'}$, is expected to be:
\begin{equation}
n_{e'}=\frac{\rho}{m_p}(1-\frac{Y_{He'}}{2}),
\label{n_e'}
\end{equation}
where $m_p$ is the mass of proton and $Y_{He'}$ is the mass fraction of $He'$, $Y_{He'}=n_{He'}m_{He}/(n_{H'}m_{H}+n_{He'}m_{He})$.

For the interaction cross section, the ordinary and mirror electron interacts through the photon-mirror photon kinetic mixing, which leads to the mirror electron holding a charge $\epsilon e$ \cite{review,cdms}. This enables the mirror electron interacts with free ordinary electron through Rutherford scatter with cross section
in the non-relativistic limit
\begin{equation}
\frac{d\sigma}{dE}=\frac{\lambda}{E^2v^2},
\label{crosssection}
\end{equation}
where
\begin{equation}
\lambda=\frac{2\pi\epsilon^2\alpha^2}{m_e}
\label{lambda}
\end{equation}
and $E$ is the recoil energy of the target electron and $v$ is the incoming electron velocity.

In principle, it would be desirable to account for the cross section of mirror electrons on bound atomic electrons which is beyond the scope of this initial study. In this work, we adopt the approximation made in \cite{cdms,prediction} that the mirror electron scatter with only the loosely bound Xe electrons with the number of $g=44$ \cite{prediction}. And we expect this approximation is valid enough for this initial study. With this approximation, the differential interaction rate is

\begin{eqnarray}
\label{differential}
\frac{dR}{dE}&=&gN_{T}n_{e'}\int \frac{d\sigma}{dE}\frac{f_{e'}(v)}{k}|v|d^3v  \nonumber\\
&=& gN_{T}n_{e'} \frac{\lambda}{E^2}\int_{|v|>v_{min}}^\infty \frac{f_{e'}(v)}{k|v|}d^3v.
\end{eqnarray}
where the lower velocity limit is $v_{min}=\sqrt{\frac{2E_R}{m_e}}$,  $N_{T}$ is the number of target Xe atoms per tonne and k=$[\pi v_0^2(e')]^{3/2}$ is the Maxwellian distribution normalization factor.

However, some recent studies have revealed that the mirror electron distribution around the earth is strongly affected by the earth-bound dark matter \cite{review,darkside50,shield}. Naive estimates would put the $e'$ capture rate in the Earth several orders or magnitude higher than mirror nuclei capture rate, which is not possible as mirror electric charge would build up and modify the $e'$ distribution, effectively shielding the detector from the halo $e'$ particles. The actual shielding mechanism is quite complex however, as mirror $E'$ and $B'$ fields can be generated not only in the Earth but also in the halo plasma near the Earth, and collisional shielding can also play a role. It has been suggested in \cite{shield} that the effect of this shielding is to not only suppress the $e'$ number density at the detector, but also to effectively provide a velocity cut-off where only the high velocity tail of the $e'$ distribution reaches the detector. 

In \cite{darkside50}, the author proposed a simple model assuming the dark electron velocity distribution with a mean speed, $\left \langle |{\bf v}|  \right \rangle  \gg v_{min}$. This condition could only be valid for electron recoil energies below some threshold, $ E_R< E_R^T$, since $v_{min} \propto E_R$. For the energy recoil greater than $E_R^T$, the scattering rate is strongly suppressed.

The integration of eq.(\ref{differential}) then becomes 
\begin{equation}
\label{intergration}
\frac{dR}{dE}=gN_{T}n_{e'} \frac{\lambda}{v_c^0 E^2_R} \quad  {\rm   for  } \ E_R<E_R^T,
\end{equation}
where $\frac{1}{v_c^0} \equiv  \int_{|v|>v_{min}}^\infty \frac{f_{e'}(v)}{k|v|}d^3v$ and $E_R^T \sim \frac{1}{2} m_e (v_c^0)^2$. Given the assumption that  $\left \langle |{\bf v}|  \right \rangle  \gg v_{min}$ for E  $ <E_R^T$, $v_c^0$ is independent on the $E_R$ here.



\section{The results}

To compare with the experimental result, we need to convolve the expression of eq.(\ref{intergration}) with the energy resolution, a Gaussian distribution with
energy-dependent width \cite{Xenon1t}
\begin{equation}
\frac{dR}{dE_m}=\frac{1}{\sigma\sqrt{2\pi}}\int\frac{dR}{dE}e^{-(E-E_m)^2/2\sigma^2}dE
\label{resolution}
\end{equation}
where $E_m$ is the measured recoil energy and $\sigma$ is the detector averaged resolution \cite{resolution} that reads
\begin{equation}
\sigma=E_m \times (\frac{31.71}{E_m}+0.15)\%, \quad {\rm For \, E_m \sim keV }.
\label{sigma}
\end{equation}

And we also need to take into account the detector efficiency, which is reported in \cite{Xenon1t}. So the final expression to be compared with the data is
\begin{equation}
\frac{dR}{dE_m}=\frac{1}{\sigma\sqrt{2\pi}}\int_{1keV}^{E_R^T}\frac{dR}{dE}e^{-(E-E_m)^2/2\sigma^2}\times \gamma(E) dE,
\label{final}
\end{equation}
where $\gamma(E)$ represents the detector efficiency.

\begin{figure}[htbp]

\subfigure[$\epsilon\alpha=1.2\times10^{-12}, \quad E^T_R=7.6keV$]{\includegraphics[width=1\columnwidth]{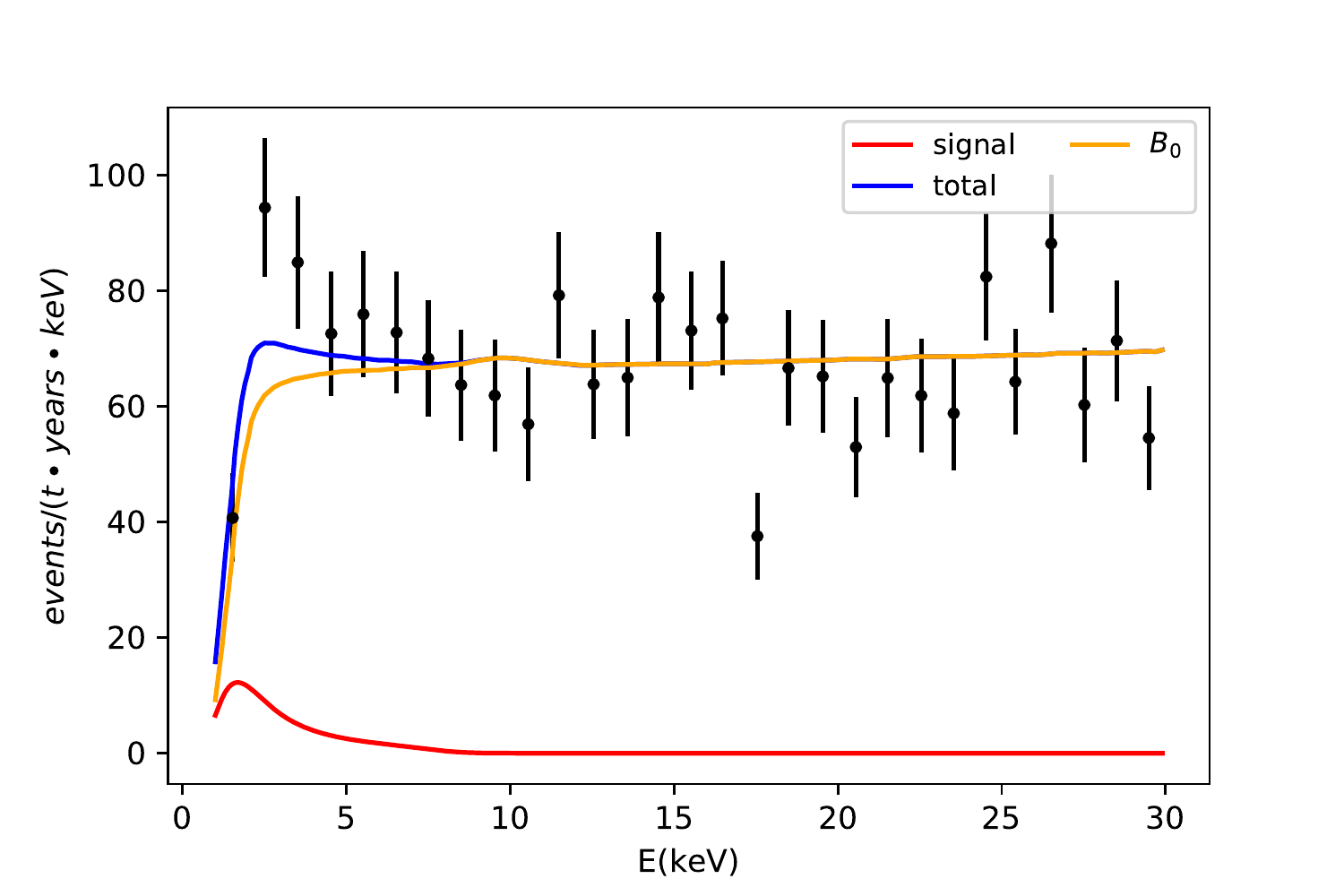}}

\subfigure[$\Delta\chi^2$]{\includegraphics[width=1\columnwidth]{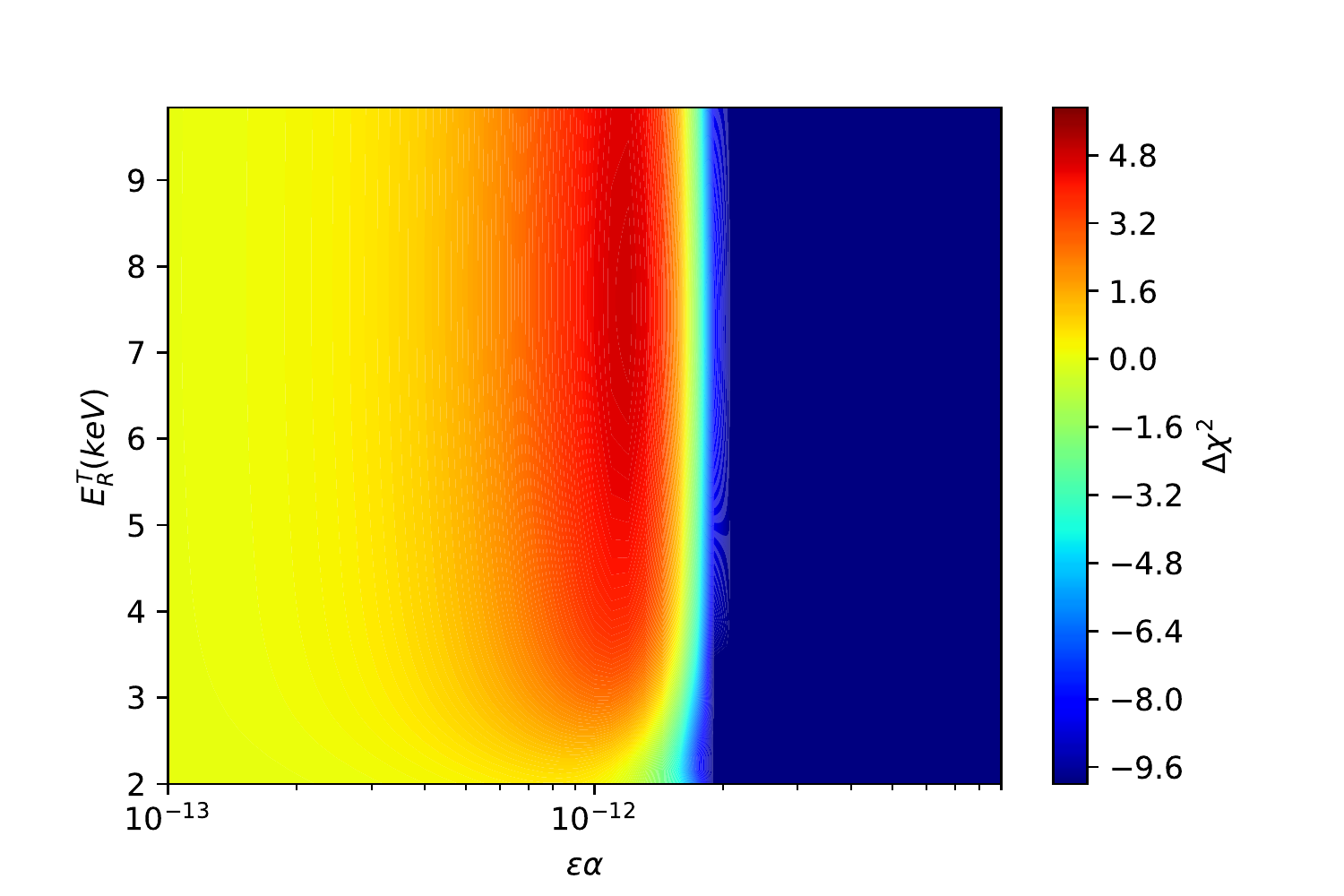}}

\caption{ Fig.(a) is the best fit result ($\Delta \chi^2=4.8, {\rm p\,value}<0.1$) with $\epsilon\alpha=1.2\times10^{-12}$ and $E^T_R=7.6keV$.  The orange line is the $B_0$ background in \cite{Xenon1t}. The red line is the signal from mirror electron scattering and the blue line is the total events predicted by the model. Fig.(b) is the $\Delta\chi^2$ contour map, where $\Delta\chi^2 = \chi^2({\rm only\, background\, B_0})-\chi^2({\rm with\, mirror\, electron\, model})$. The x-axis is the $\epsilon\alpha$ and the y-axis is the $E_R^T$.}

\label{result}

\end{figure}

With the definition $\alpha=\sqrt{\frac{n_{e'}/0.2cm^{-3}}{v_c^0/50000km/s}}$, we have $\frac{dR}{dE} \propto (\epsilon\alpha)^2$ according to eq.(\ref{intergration}). Thus there are two free parameters, $E^T_R$ and $\epsilon\alpha$, for the scattering rate. The fit results are shown in Fig.\ref{result}. The best fit result ($\Delta \chi^2=4.8, {\rm p \, value}<0.1$) with $\epsilon\alpha=1.2\times10^{-12}$ and $E^T_R=7.6keV$ is shown in Fig.(a). The signal from mirror electron scattering rises below 5keV and peaks at about 2keV, which has been predicted in \cite{prediction}. The $\Delta \chi^2 \equiv\chi^2({\rm only\, background\, B_0})-\chi^2({\rm with\, mirror\, electron\, model})$ contour map is shown in Fig.(b). Fitting result is highly correlated to $\epsilon\alpha$ and weakly to $E^T_R$ for $E^T_R > 4keV$. Mirror electron model with $\epsilon\alpha \sim 10^{-12}$ and $E_R^T \geq 4{\rm keV}$ provides a good fit for the low energy recoil spectrum. 

\begin{figure}[htbp]
\subfigure{\includegraphics[width=1\columnwidth]{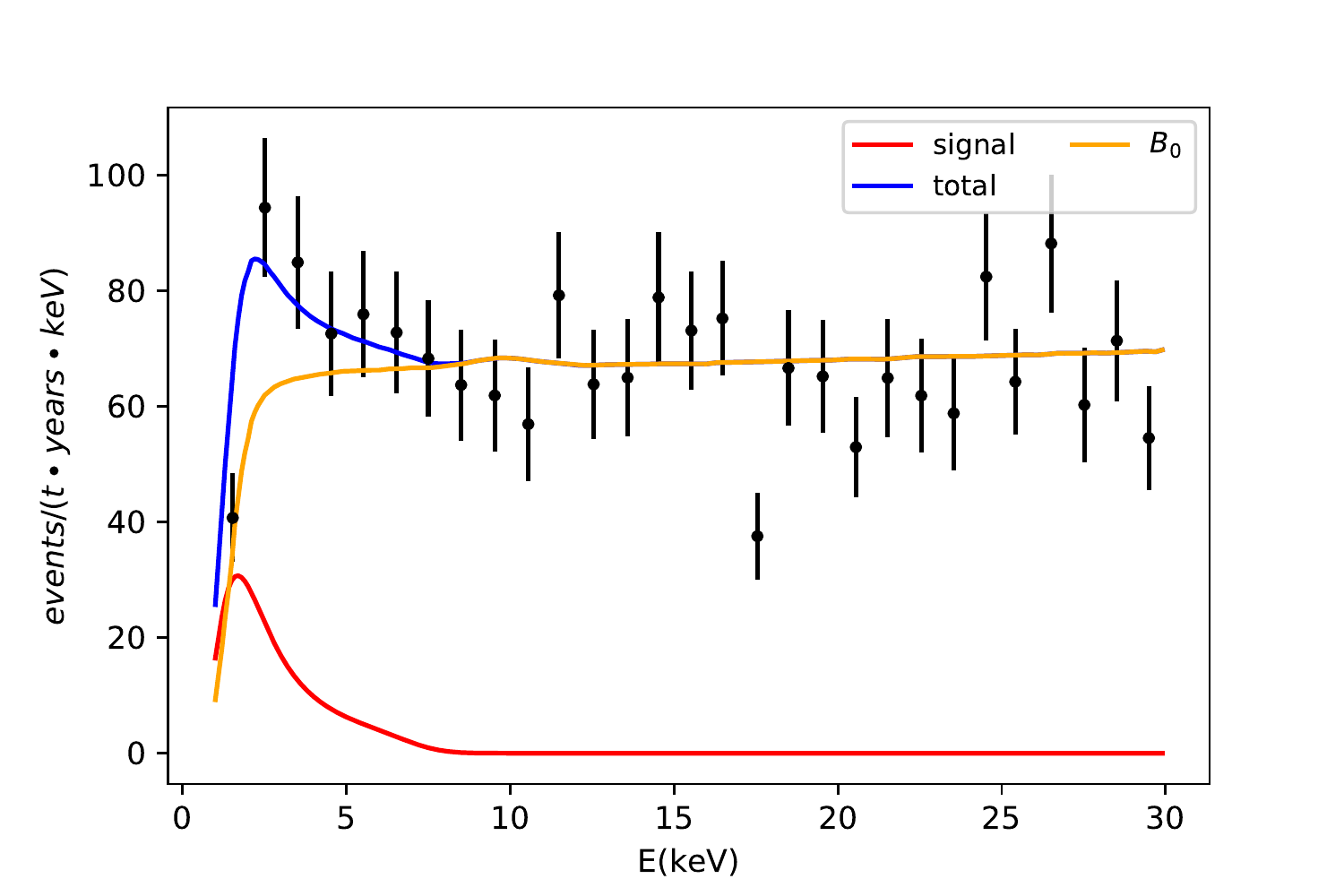}}

\caption{The fit result with $E_R^T=7keV$ and $\epsilon\alpha = 1.9 \times 10^{-12}$.}

\label{peak}

\end{figure}

It is also interesting to note that the mirror electron model can also fit the peak around 2keV but at the cost of the error of first data point ($E_R \sim 1 {\rm keV}$), which leads to a higher $\Delta \chi^2$. Such fit result is shown in Fig.\ref{peak} for example.

\begin{figure}[htbp]
\subfigure[$\epsilon\alpha=9.1\times10^{-13}$, $E_R^T=6.6keV$]{\includegraphics[width=1\columnwidth]{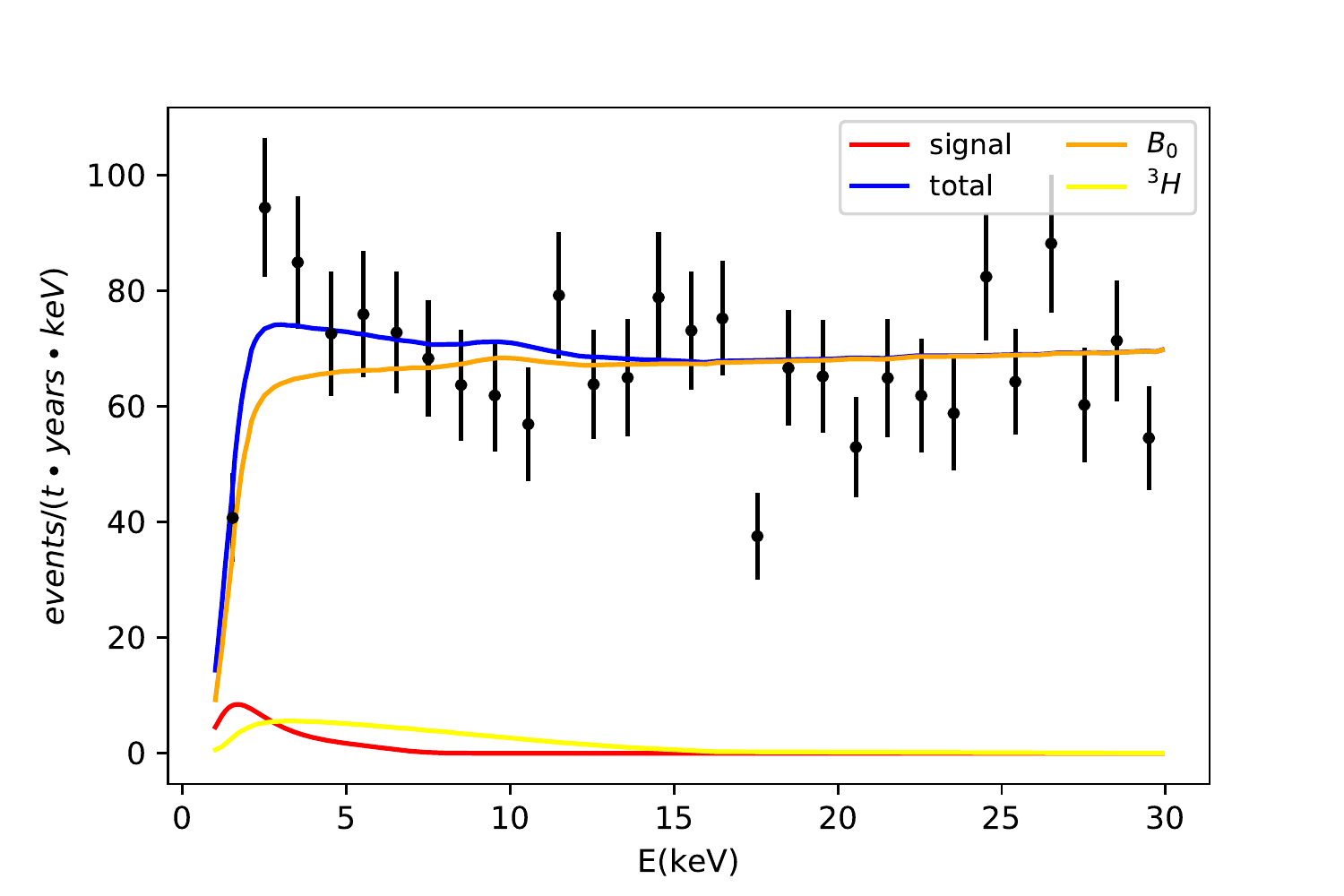}}
\subfigure[$\Delta \chi^{2}$]{\includegraphics[width=1\columnwidth]{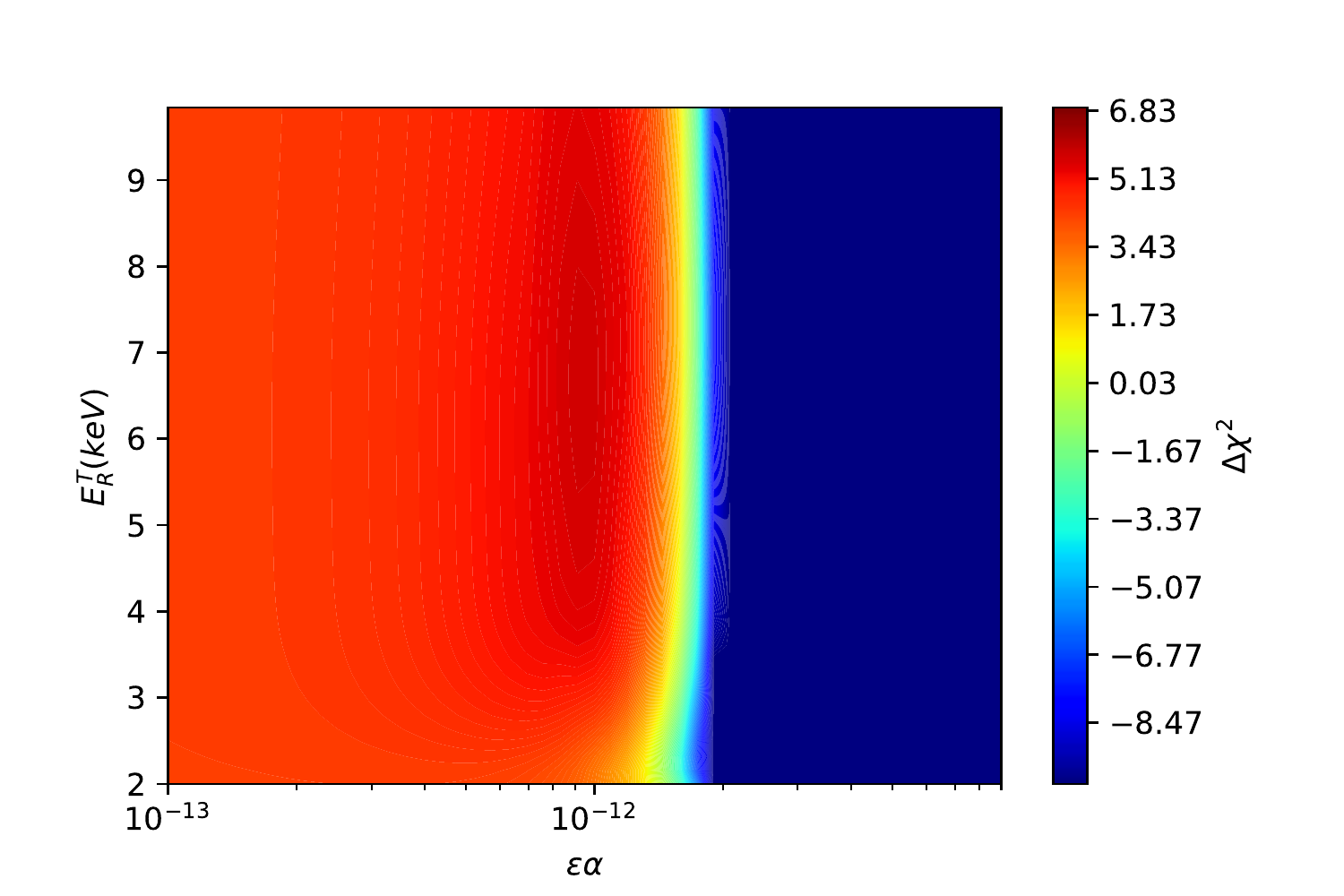}}
\caption{Fig.(a) is the best fit result ($\Delta \chi^2 =5.7$) including an unconstrained tritium component with $ \epsilon\alpha=9.1 \times 10^{-13}$ and $ E_R^T=6.6keV $. The yellow line represents the $^3H$ component. Fig.(b) is the $\Delta\chi^2 = \chi^2({\rm only\, background\, B_0})-\chi^2({\rm with \, ^3H\, and \, mirror\, electron\, model})$ contour map, but with an unconstrained $^3H$ component.}

\label{h3}

\end{figure}

Considering the limited understanding on the $^3H$ background \cite{Xenon1t}, we fit the data with an unconstrained tritium component. The best fit result and the $\Delta\chi^2$ contour map is shown in Fig.\ref{h3}. The tritium component signal is comparable with the mirror electron scattering for the best fit result.

{Our result is well consistent with previous direct detection results like Darkside50 ($\epsilon\alpha \leq 1.5 \times 10^{-11}$\cite{darkside50}) and LUX ($\epsilon \leq 10^{-11}$ at $\rm{T \sim 0.3keV}$ \cite{LUX}). It has been argued that mirror dark matter with $\epsilon \sim 7 \times 10^{-10}$ could account for the CDMS low energy electron recoil spectrum \cite{cdms}. However, that result is in tension with the null result in LUX \cite{LUX}. In \cite{cdms} the author has not considered the shielding effect, which is actually not negligible in the direct detection according to the following studies\cite{shield}. This shielding effect makes our results in the form like \cite{darkside50} but different with \cite{cdms}. The kinetic mixing constant $\epsilon \sim 10^{-10}$ is suggested by small scale structure and the mirror big bang nucleosynthesis  \cite{review}. With the $\epsilon\alpha \sim 10^{-12}$ supported by the XENON1T low energy electron recoil spectrum, $\epsilon \sim 10^{-10}$ suggests $\alpha \sim 10^{-2}$, which means that either $v_c^0 \gg 50000km/s$ and/or $n_{e'} \ll 0.2cm^{-3}$. Similar results have also been shown in other electron recoil direct detection \cite{shield,darkside50}. When mirror dark matter is captured and accumulates within the Earth, it forms an extended distribution and leading to the formation of a ‘dark ionosphere’. This shielding effect provide a mechanism for reducing $n_{e'}$, and suppress the flux of halo dark matter particles below some cutoff, $v_{cut}$ \cite{shield}. And we will focus on this issue in our future work.


\section{conclusion}

In this work, we suggest that the mirror dark matter model may account for the low energy electronic recoil spectrum in XENON1T. The allowing parameter space ($\epsilon\alpha \sim 10^{-12}$ and $E_R^{T}$) is within the limits of Darkside50. In 2020, several experiments such as XENONnT\cite{xenonnt}, LUX-ZEPLIN\cite{Lux}, PandaX-4T\cite{pandax}, will start their performance. In a few years, the electronic recoil spectrum, in particular in the low energy range, will be independently measured by these experiments with much higher accuracy. The mirror dark matter interpretation for the current potential excess will be stringently tested.

{\bf Acknowledgments}: We thank R.Foot for theoretical discussion. This work is supported by the National Key Research
and Development Program of China (Grant No. 2016YFA0400200), the National
Natural Science Foundation of China (Grants No. 11525313,
No. 11773075, No. U1738210, No. U1738136, and No. U1738206), the 100
Talents Program of Chinese Academy of Sciences.

\end{document}